\documentclass[10pt]{article}
\usepackage{fullpage}
\usepackage{amsmath}
\usepackage{amssymb}
\usepackage[dvips]{epsfig}
\usepackage{color}
\usepackage{subfigure}

\def\##1{{\bf #1}}
\def\-#1{\underline #1}
\def\=#1{\underline{\underline #1}}

\def\eps{\epsilon}

\def\ko{k_0}

\def\.{\mbox{ \tiny{$^\bullet$} }}
\def\etaL{\eta/L}
\def\koL{k_{0}L}

\def\ux{\#{u}_x}
\def\uy{\#{u}_y}
\def\uz{\#{u}_z}

\def\un{\#{u}_n}
\def\ut{\#{u}_\tau}
\def\ub{\#{u}_b}

\def\le{\left(}
\def\ri{\right)}
\def\les{\left[}
\def\ris{\right]}
\def\lec{\left\{}
\def\ric{\right\}}

\def\c#1{\cite{#1}}
\def\l#1{\label{#1}}
\def\r#1{(\ref{#1})}

\begin{document}

\begin{center}

{\bf {\LARGE On columnar thin films as platforms for
surface-plasmonic-polaritonic optical sensing: higher-order
considerations}}

\vspace{10mm}

 {\bf {\large Siti S. Jamaian${}^{a}$ and Tom G. Mackay${}^{a,b,}$\footnote{E--mail: T.Mackay@ed.ac.uk.}}}\\

\vspace{5mm}

${}^{a}$School of Mathematics and
   Maxwell Institute for Mathematical Sciences, University of Edinburgh, Edinburgh EH9 3JZ, UK\\
${}^{b}$NanoMM~---~Nanoengineered Metamaterials Group\\ Department of Engineering Science and Mechanics,
Pennsylvania State University,\\
University Park, PA 16802--6812, USA\\

\normalsize

\vspace{15mm}
{\bf Abstract}

\end{center}

\vspace{4mm}

The ability to tailor the porosity and optical properties of
columnar thin films (CTFs) renders them promising platforms for
optical sensing. In particular, surface-plasmon-polariton (SPP)
waves, guided by the planar interface of an infiltrated CTF and  a
thin layer of metal, may be harnessed to detect substances that
penetrate the void regions in between the columns of a CTF. This
scenario was investigated theoretically using a higher-order
homogenization technique, based on an extended version of the
second-order strong-permittivity-fluctuation theory, which takes
into account the size of the component particles which make up the
infiltrated CTF and the statistical distribution of these particles.
Our numerical studies revealed that as the size of the component
particles increases and as the correlation length that characterizes
their distribution increases: (i)  the phase speed of the SPP wave
decreases and the SPP wave's attenuation increases; (ii) the SPP
wave's penetration into the CTF decreases; (iii) the angle of
incidence required to excite the SPP wave in a modified Kretschmann
configuration increases; (iv) the sharpness of the SPP trough in the
graph of reflectance versus angle of incidence increases;  and (v)
the sensitivity  to changes in refractive index of the infiltrating
fluid decreases.

 \vspace{5mm}


\section{Introduction}

A columnar thin film (CTF) consists of a parallel array of columnar
nanowires, oriented at angle $\chi$ to a planar substrate
\c{STF_Book}. The columns may be grown on the substrate by means of  physical vapour
deposition. Through judicious control of the vapour deposition
process,  both the macroscopic optical properties and the porosity
of the CTF can be tailored to order \c{Smith_AO}. Consequently, CTFs
are promising candidates as platforms for optical sensing
applications, wherein it is envisaged that the species to be sensed
is contained within a fluid which penetrates the void regions in
between the CTF's columns.

A recent theoretical investigation highlighted the fact  that
surface-plasmon-polariton (SPP) waves, guided by the planar
interface of an infiltrated CTF and  a thin layer of metal, may be
usefully harnessed for optical sensing \c{ML_PNFA}. In this
scenario, the angle of incidence required to excite an SPP wave was
found to be acutely sensitive to changes in refractive index
$n_\ell$ of the fluid which fills the CTF's void regions.
Furthermore, the phase speed and propagation length of the SPP wave
were also sensitive to $n_\ell$.

A crucial step in modelling CTFs~---~as well as more general
sculptured thin films~---~as platforms for such optical sensing is
the estimation of the effective permittivity dyadic of the
infiltrated CTF \c{Sherwin}. This may be achieved in a two-step
homogenization process: First, an inverse homogenization procedure
can be implemented to estimate  certain nanoscale parameters
(namely, refractive index, shape parameter and volume fraction) for
the uninfiltrated CTF, using experimentally-determined values of the
uninfiltrated CTF's permittivity dyadic \c{ML_JNP}. Second, a
forward homogenization approach can then be used to estimate the
effective permittivity dyadic of the infiltrated CTF. In recent
studies the well-established Bruggeman homogenization formalism was
employed for this purpose \c{ML_PNFA,ML_PJ,ML_IEEE_SJ}. However,
there are certain limitations with the Bruggeman formalism.
 It does not take into account the nonzero size of
particles which make up the component materials nor does  it take
into account the statistical distribution of these component
particles. But both the size and distribution of the component
particles can have a significant bearing upon the estimates provided
by homogenization formalisms. For example, significant higher-order
effects associated with the nonzero size of component particles were
recently reported for a homogenization study based on an infiltrated
chiral sculptured thin film \c{JM_OC}.

In the following we investigate CTFs as platforms for SPP optical
sensing  using a higher-order approach to homogenization, based on
the strong-permittivity-fluctuation theory (SPFT) \c{TK81}. A
second-order implementation is used wherein the statistical
distributions of the component particles are characterized in terms
of a two-point correlation function and its correlation length $L$
\c{Genchev,Z94,MLW01}. Also, an extended implementation is used
wherein the nonzero size of component particles is accommodated via
a size parameter $\eta$ which originates from the corresponding
depolarization dyadics \c{M_WRM}. Two SPP scenarios are
investigated: (i) a canonical scenario wherein the SPP wave is
guided by the planar interface of a CTF half-space and a metal
half-space (Sec.~\ref{SWP}); and (ii) a realistic scenario for SPP
wave excitation based on a modification of the Kretschmann
configuration (Sec.~\ref{MKC}). Our particular goal is to elucidate
the effects of the higher-order homogenization parameters $L$ and
$\eta$ on SPP wave excitation. As the theory which underpins our
study is  described comprehensively elsewhere, we provide only a
brief outline of the theory (along with appropriate citations to the
recent literature) before presenting our numerical results.

As regards notation, vectors are written in boldface while
3$\times$3 dyadics are underlined twice. The unit vectors $\#u_x$,
$\#u_y$ and $\#u_z$ are aligned with the Cartesian $x$, $y$ and $z$
axes. The real and imaginary parts of complex quantities are
provided by the operators $\mbox{Re}\les \cdot \ris$ and
$\mbox{Im}\les \cdot \ris$; and $i = \sqrt{-1}$. An $\exp \le - i
\omega t \ri$ time-dependence is implicitly assumed, with $\omega$
being the angular frequency and $t$ being time. The free-space
wavenumber is denoted as $\ko$.

\section{Homogenization}\label{theory}

\subsection{Macroscopic perspective}

Macroscopically, a CTF may be regarded as a homogeneous biaxial
dielectric material, characterized by a relative permittivity dyadic
of the form \c{STF_Book,Smith_AO}
\begin{equation} \l{dyadic_eps}
\=\eps_{\,ctf\nu} = \eps_{a\nu}  \,\un\un +\eps_{b\nu}\,\ut\ut \,
+\,\eps_{c\nu}\,\ub\ub, \qquad \quad \le \nu = 1, 2 \ri.
\end{equation}
Herein $\nu =1 $ denotes an uninfiltrated CTF and $\nu=2$ an
infiltrated CTF, while the  normal, tangential and binormal basis
vectors are given by
\begin{equation}
\left. \begin{array}{l}
 \un = - \ux \, \sin \chi + \uz \, \cos \chi \vspace{4pt} \\
 \ut =  \ux \, \cos \chi + \uz \, \sin \chi \vspace{4pt} \\
\ub = - \uy
\end{array}
\right\}
\end{equation}
with the column inclination angle $\chi \in \le 0, \pi/2 \ris$. The
 permittivity parameters for certain uninfiltrated CTFs have been
experimentally determined. For example,
\begin{equation}
\left.
\begin{array}{l}
\eps_{a1} = \displaystyle{\les 1.0443 + 2.7394 \le \frac{2
\chi_v}{\pi} \ri - 1.3697
\le \frac{2 \chi_v}{\pi} \ri^2 \ris^2} \vspace{6pt} \\
\eps_{b1} = \displaystyle{ \les 1.6765 + 1.5649 \le \frac{2
\chi_v}{\pi} \ri - 0.7825 \le \frac{2 \chi_v}{\pi} \ri^2 \ris^2}
 \vspace{6pt} \\
\eps_{c1} = \displaystyle{ \les 1.3586 + 2.1109 \le \frac{2
\chi_v}{\pi} \ri - 1.0554 \le \frac{2 \chi_v}{\pi} \ri^2 \ris^2}
\end{array}
\right\} \l{tio1}
\end{equation}
for a
 CTF made from
patinal${}^{\mbox{\textregistered}}$  titanium oxide, as determined
at a free--space wavelength of 633 nm \c{HWH_AO}. Here $\chi_v$ is
the angle of incidence of the vapour flux during the deposition
process used to fabricate the CTF. It  is related to the column
inclination angle $\chi$ via
\begin{equation}
\tan \chi = 2.8818 \, \tan \chi_v. \l{tio2}
\end{equation}
For definiteness, we fix $\chi_v = 30^\circ$.
 From a knowledge
of $\lec \eps_{a1}, \eps_{b1}, \eps_{c1} \ric$, the corresponding
permittivity parameters for a CTF infiltrated by a fluid of
refractive index $n_\ell$, namely  $\lec \eps_{a2}, \eps_{b2},
\eps_{c2} \ric$, can be estimated using a inverse/forward
homogenization process, as  we now describe.

\subsection{Nanostructural perspective}

Within our homogenization approach, the columns of the CTF are
envisaged at the nanoscale as assemblies of ellipsoidal particles,
all with their major axes aligned with the rotational axis of the
columns. Thus, the surface of each ellipsoidal particle, relative to
its centre, is prescribed by the position vector
\begin{equation} \l{U_def}
 \eta \, \=U \cdot \#{\hat r} ,
\end{equation}
where  $\#{\hat r}$ is the radial unit vector which prescribes the
surface of the unit sphere. In conformity with eq.~\r{dyadic_eps},
the  shape dyadic $\=U$ has the form
\begin{equation}
\=U = \frac{1}{\sqrt{\gamma_\tau \gamma_b }} \le \un \, \un +
\gamma_\tau \, \ut \, \ut + \gamma_b \, \ub \, \ub \ri .
\end{equation}
A suitably elongated ellipsoidal shape is achieved by choosing the
shape parameters $\gamma_{b} \gtrsim 1$ and $\gamma_\tau \gg 1$. The
value $\gamma_\tau = 15$ is selected here because larger values of
$\gamma_\tau$ do not result in any significant effects for slender
inclusions \c{Lakh_Opt}. The size parameter $\eta$ provides a
  measure of the linear  dimensions of the ellipsoidal particle.
  In compliance with
the homogenization regime, $\eta$ must be smaller than the
wavelengths involved, but it need not be vanishingly small. In
conventional homogenization formalisms, as typified by the Bruggeman
formalism, the limit $\eta \to 0$ is taken, whereas in extended
formalisms nonzero values of  $\eta$ are incorporated \c{ML_PiO}.

The  regions in between the columns of the CTFs (whether
infiltrated or not) are taken to have  the same particulate
nanonstructure as the columns themselves. So that there are no
spaces in between the ellipsoidal component particles,  a
fractal-like distribution of the component particles is envisaged.
Accordingly, the size parameter $\eta$ should be regarded as an
upper bound on the linear dimensions of the component particles
\c{M_JNP}.

 In
conventional homogenization formalisms, as typified by the Bruggeman
formalism, the distributions of the component materials are
characterized solely in terms of volume fraction: therein,
 the volume fraction of a CTF
occupied by columns is  $f \in \le 0, 1 \ri $, while $1 -f$ is the
volume fraction not occupied by columns. In addition to volume
fraction, the second-order SPFT also characterises the distribution
of the component materials in terms of a two-point correlation
function and its associated correlation length $L$ \c{TK81,MLW01}. The wavelengths
involved are  presumed to be large in relation to the correlation
length. Furthermore, on the grounds of physical reasonableness, the
inequality $L > \eta$ should be satisfied \c{M_WRM}.

The deposited material from which the columns of the CTF are
composed is assumed to be an isotropic dielectric material of
refractive index $n_s$. This refractive index is generally somewhat
different to the refractive index of the bulk material that was
evaporated to make the CTF, depending upon the precise details of
the deposition environment \cite{WRL03,Macleod}.

We emphasize that the nanoscale parameters $\lec \gamma_b, f, n_s
\ric$ are not readily determined by direct measurement. Therefore,
we rely upon  theoretical estimates of these quantities.

\subsection{Inverse homogenization~---~uninfiltrated CTF} \l{InverseHomog}

The first step towards the estimation of the infiltrated CTF's
permittivity parameters $\lec \eps_{a2}, \eps_{b2}, \eps_{c2} \ric$
involves an application of inverse homogenization. The (nonextended)
Bruggeman formalism is inverted to yield estimates of the shape
parameter $\gamma_b$, volume fraction $f$ and refractive index
$n_s$, making use of   the experimentally-determined values of $\lec
\eps_{a1}, \eps_{b1}, \eps_{c1} \ric$ provided in eqs.~\r{tio1} for
the uninfiltrated CTF. This procedure is described in detail
elsewhere \c{ML_JNP}. Parenthetically,  certain constitutive
parameter regimes are known to be problematic for the  inverse
Bruggeman formalism \c{SSJ_Inv_Br} (and, indeed,  also problematic
for the forward Bruggeman formalism \c{ML_Br}) but these regimes are
not the same as those considered here.

\subsection{Forward homogenization~---~infiltrated CTF}\l{2ndSPFT}

The second step towards the estimation of the infiltrated CTF's
permittivity parameters $\lec \eps_{a2}, \eps_{b2}, \eps_{c2} \ric$
involves combining the estimated nanoscale parameters $\lec
\gamma_b, f, n_s \ric$  with the refractive index of the
infiltrating fluid $n_\ell$ in an application of forward
homogenization. In contrast with recent studies, we implement an
extended version of the second-order SPFT for this purpose. Thereby,
the distributional statistics of the component materials as well as
the nonzero size of the component particles are taken into account.
A full description of the extended version of the second-order
SPFT~---~appropriate to biaxial homogenized composite materials,
based on oriented ellipsoidal component particles, precisely as is
the case here~---~is available elsewhere \c{M_WRM,M_JNP}. Let us
note that in the limits $L \to 0$ and $\eta \to 0$,  the
nonextended, zeroth-order SPFT emerges which is equivalent to the
Bruggeman formalism. Also,
  higher-order homogenization approaches such as those implemented here generally
do not scale in a physically-intuitive manner \c{Bohren_scaling},
but
 the physical significance of this aspect is unclear.

Numerical estimates of $\eps_{a2}$,  $\eps_{b2}$ and
 $\eps_{c2}$, as computed using the extended version of the second-order SPFT,
  are plotted against $\eta/L \in (0,1)$ and $k_{0}L \in (0,0.2)$ in Fig.~\ref{Figure1}.
To allow direct comparison with an earlier study based on the
Bruggeman formalism \c{ML_PNFA},  the refractive index of the fluid
infiltrating the void regions of the CTF was fixed at $n_\ell =
1.5$. The real parts of $\lec \eps_{a2}, \eps_{b2}, \eps_{c2} \ric$
increase modestly, but significantly,  as both $L$ and $\eta$
increase. More conspicuous is the uniform increase in the imaginary
parts of $\lec \eps_{a2}, \eps_{b2}, \eps_{c2} \ric$ from the null
values that they take in the limits $L \to 0$ and $\eta \to 0$. The
emergence of these nonzero imaginary parts is indicative of
 losses due to scattering from the macroscopic
coherent field \c{Kranendonk}. These scattering losses increase as
the component particles increase in size and as the correlation
length increases. Such scattering losses  are a general feature of
higher-order approaches to homogenization, arising whenever  the
component materials are characterized by  a (nonzero) length scale
 \c{Doyle,Prinkey}.

\section{Canonical boundary-value problem} \l{SWP}

Now we turn to the excitation of SPP waves, guided by the planar
interface of a CTF half-space and metal half-space. This  scenario
allows us to explore the wavenumbers of the SPP waves and thereby
consider the phase speeds and propagation lengths of the SPP waves.
The theory for this canonical boundary-value problem   was developed
recently by Lakhtakia and Polo \c{LP_AJP}.

Let assume that a metal of relative permittivity $\eps_m$  occupies
the half-space $z< 0$ while a CTF infiltrated by a fluid of
refractive index $n_\ell$ occupies the half-space $z> 0$. We
consider the $p$-polarized SPP wave which is guided by the planar
interface. The electric  field phasor of the SPP wave may be
expressed as \c{LP_AJP}
\begin{equation}
\#E (\#r) = \left\{ \begin{array}{lcr}  \displaystyle{A_m \le \ux -
\frac{i \sigma}{q_m} \uz \ri \, \exp \les i \ko \le \sigma x - i q_m
z
\ri \ris} , && z < 0 \vspace{6pt} \\
 \displaystyle{A_c \les \ux + \frac{i \sigma q_c - \le \eps_{a2} -
\eps_{b2} \ri \, \sin \chi \, \cos \chi}{\sigma^2 - \le \eps_{a2}
\cos^2 \chi + \eps_{b2} \sin^2 \chi \ri}\, \uz \ris \, \exp \les i
\ko \le \sigma x + i q_c z \ri \ris}, && z > 0
\end{array} \right.,
\end{equation}
where $A_m$ and $A_c$ are the complex--valued amplitudes, $\sigma
\ko \ux$ represents the wave vector of the SPP wave, and
 $q_m =
\sqrt{\sigma^2 - \eps_m}$. The relative wavenumber $\sigma$ and
$q_c$ are deduced from  the corresponding dispersion relations. The
roots of the dispersion relations are chosen such that $\mbox{Re}
\,\les \, q_{m,c} \, \ris > 0$, thereby ensuring that the
 SPP  wave decays in directions normal to the interface $z=0$.

For our numerical studies we
 chose $\eps_m = -56 + 21i$ (which is the value for bulk aluminium at a free--space
wavelength of $633$ nm), in keeping with previous studies \c{ML_PNFA}.
 In Fig.~\ref{Figure2}, the real and imaginary
parts of $\sigma$ are plotted against $\eta/L \in (0,1)$ and $k_{0}L
\in (0,0.2)$.
 The real part of $\sigma$ increases uniformly as both the correlation length
and the size parameter increase. Accordingly,  a significant
decrease in the phase speed of the SPP wave may be inferred  as both
$L$ increases and $\eta$ increases. As the  imaginary part of
$\sigma$ similarly increases in a uniform fashion as both the
correlation length and the size parameter increase, we deduce that
attenuation of the SPP wave is significantly increased upon
increasing  both $L$ and $\eta$.

Next we turn to the penetration of the SPP wave into the CTF in the
$+z$ direction. An inverse measure of this is provided by $\mbox{Re}
\les  q_c \ris$, which is
 plotted
against $\eta/L$ and $k_{0}L$  in Fig.~\ref{Figure3}. Thus, from
Fig.~\ref{Figure3} we deduce that SPP wave penetration into the CTF
decreases as the correlation length increases and as the size
parameter increases.

\section{Modified Kretschmann configuration}\label{MKC}

The Kretschmann configuration \c{Kret}  represents a realistic
scenario for SPP wave excitation. A modification of this
configuration is considered here \c{Lakh_oc}, guided by experimental
views for launching surface waves \c{Simon_1975},  wherein a
fluid--infiltrated CTF of finite thickness occupies the region $L_m
< z < L_\Sigma$ and a thin metal film occupies the region $0 < z <
L_m$. The region $z > L_\Sigma$ is filled by a dielectric material
with relative permittivity   $\eps_{\ell}=n_\ell^2$, while the
region $z < 0$ is filled by a dielectric material with relative
permittivity   $\eps_{d}$.

Suppose that
 a $p$--polarized plane wave propagates in the region $z \leq 0$
towards the metal--coated CTF at an angle  $\theta_{inc} \in \le 0,
\pi/2 \ri$ to the $+z$ axis. This incident plane wave gives rise to a
reflected plane wave in the $z < 0$ region and a transmitted plane
wave in the $z > L_\Sigma$ region. Thus, the corresponding electric
field phasors in the regions above and below the  metal--coated CTF
may be expressed as \c{ML_PNFA,LP_AJP}
\begin{equation}
\#E = \left\{
\begin{array}{lcr}
 \le - \ux \cos \theta_{inc} + \uz \sin \theta_{inc} \ri \,
\exp \les i \le \kappa x + \sqrt{\eps_{d}}\, z \cos \theta_{inc} \ri
\ris && \vspace{3pt} \\ +  r_p  \le  \ux \cos \theta_{inc} + \uz
\sin \theta_{inc} \ri \, \exp \les i \le \kappa x -
\sqrt{\eps_{d}}\, z \cos \theta_{inc}
\ri \ris, && z < 0  \vspace{6pt}\\
 t_p  \le - \ux \cos \theta_{tr} + \uz \sin \theta_{tr}
\ri \, \exp \les i \le \kappa x + \sqrt{\eps_{\ell}}\,\le z -
L_\Sigma \ri \, \cos \theta_{tr} \ri \ris, && z > L_\Sigma
\end{array}
\right.,
\end{equation}
wherein the angles of incidence and transmission are related by
\begin{equation}
\sqrt{\eps_{d}} \, \sin \theta_{inc} = \sqrt{\eps_{\ell}} \, \sin
\theta_{tr} \equiv \sigma
\end{equation}
and $\kappa = \ko \sigma$. By solving the associated boundary
value-problem, the complex-valued reflection and transmission
coefficients, namely $r_p$ and $r_t$, are determined.

We may identify the excitation of a SPP wave at the metal--CTF
interface via the reflectance $|\, r_p \, |^2$ or the absorbance
\begin{equation}
A_p = 1 - \le  |\, r_p \, |^2 + \beta\, | \, t_p \, |^2 \ri,
\end{equation}
wherein the scalar parameter
\begin{equation}
\beta = \frac{\sqrt{\eps_{\ell}} \; \mbox{Re} \, \les \, \cos
\theta_{tr} \, \ris}{\sqrt{\eps_{d}} \, \cos \theta_{inc}}.
\end{equation}
A sharp trough in graph of $|\, r_p \, |^2$ (or sharp peak in the
graph of $A_p$) versus $\theta_{inc}$ is the characteristic
signature of SPP-wave excitation, provided that $\theta_{inc}$ is
greater than the critical angle for total reflection in the absence
of the metal film. In order to ensure the existence of this critical
angle, in our numerical studies we chose $\eps_{d} = 6.76$, which is
in the range of rutile.

A representative plot of the reflectance $|\, r_p \, |^2$ versus
$\theta_{inc}$ is presented in Fig.~\ref{Figure4} for $\koL = 0.2$
and $\etaL = 1$. We set the thicknesses
 $L_m = 10$
nm and $L_\Sigma = L_m + 1000$ nm. Also
 plotted in this figure is  $\beta | t_p |^2$. In the regime
 $\theta_{inc} > 36^\circ$ (which corresponds to the total reflection
  regime in the absence of the metal film),
  the value of the absorbance $A_p$ can be inferred from that of the
  reflectance $|\, r_p \, |^2$ since $\beta | t_p |^2$ is
  null-valued here.
The rightmost local minimum in the graph of $|\, r_p \, |^2$ (at
$\theta_{inc} =  70.8^\circ$) indicates the excitation of a SPP
wave. The other local minimums in the graph of $|\, r_p \, |^2$,
which occur for $36^\circ < \theta_{inc} <  70.8^\circ$, represent
waveguide modes \cite{Motyka1}. Crucially, these waveguide  modes
are excited at different values $\theta_{inc}$ as the thickness of
the CTF is varied, whereas the value of $\theta_{inc}$ which excites
the SPP wave is independent of the CTF's thickness (as we confirmed
in numerical calculations not presented here).

Let us now focus on $\theta^\sharp_{inc}$ which denotes the value of
$\theta_{inc}$  corresponding to the rightmost local minimum in the
graph of $|\, r_p \, |^2$ versus $\theta_{inc}$; i.e., it
corresponds to the value of $\theta_{inc}$  at which the SPP wave is
excited.
 The sensitivity of this angle  $\theta^\sharp_{inc}$
 to the correlation length and size parameter is
demonstrated  in Fig.~\ref{Figure6}, where $\theta^\sharp_{inc}$
is plotted against $\eta/L \in (0,1)$ and $k_{0}L \in (0,0.2)$. We
see that $\theta^\sharp_{inc}$ increases uniformly as $L$ increases
and as $\eta$ increases. In particular, the value of
$\theta^\sharp_{inc}$ for $L = 0.2 / \ko$ and $\eta = L$ is
approximately one degree greater than it is in the limits $L \to 0$
and $\eta \to 0$.

From the point of view of sensing applications, the shape of the
local minimum at  $\theta_{inc} = \theta^\sharp_{inc}$ is an
important consideration. To explore the effects of the correlation
length and size parameter on the shape of the SPP trough, the second
derivative $d^2 \le \left| r_p \right|^2 \ri / d \theta^2_{inc}$ is
 plotted against $\eta/L \in
(0,1)$ and $k_{0}L \in (0,0.2)$ in Fig.~\ref{Figure7}. We observe
that the SPP trough becomes uniformly sharper as $L$ increases, but
changes relatively little as $\eta$ increases. In particular, the
SPP trough for $L = 0.2 / \ko$ and $\eta = L$ is approximately 2\%
sharper than it is in the limits $L \to 0$ and $\eta \to 0$.

Lastly, we turn to  the sensitivity of the SPP trough in the graph
of $| r_p |^2$ versus $\theta_{inc}$ to changes in the refractive
index $n_\ell$ of the fluid which infiltrates the CTF. A convenient
measure of this sensitivity is provided by the figure of merit
 (in degree/RIU\footnote{RIU = refractive-index
unit}) \c{ML_PNFA}
\begin{equation}
\rho = \frac{\theta^\sharp_{inc} (n_{\ell2}) - \theta^\sharp_{inc}
(n_{\ell1})}{n_{\ell2} - n_{\ell1}},
\end{equation}
where $\theta^\sharp_{inc}$ is regarded as function of $n_\ell$. For
$n_{\ell2} = 1.5$ and $n_{\ell1} = 1.0$ the figure of merit is
plotted against $\eta/L \in (0,1)$ and $k_{0}L \in (0,0.2)$ in
Fig.~\ref{Figure8}. The sensitivity, as gauged by $\rho$, decreases
uniformly as both the correlation length increases and as the size
parameter increases. In particular, the sensitivity drops by
approximately 4 degree/RIU as
 $L $ and $\eta$ both increase from zero to   $ 0.2 /
\ko$.

\section{Closing remarks} \l{close}

 The potential that infiltrated CTFs offer for
SPP-based optical sensing has been  demonstrated in previous
theoretical studies using the  Bruggeman homogenization formalism.
We have further elucidated this matter by implementing a more
sophisticated homogenization approach based on the extended
second-order SPFT which takes into account the statistical
distribution of the component materials and the size of the
component materials. Specifically, our numerical investigations have
revealed that as the correlation length increases and as the size
parameter increases:
\begin{itemize}
\item the phase speed of the SPP wave decreases and the SPP wave's  attenuation
increases;
\item the SPP wave's penetration into the CTF decreases;
\item the angle of incidence required to excite the SPP wave in a
modified Kretschmann configuration increases;
\item the sharpness of the SPP trough in the graph of reflectance versus
angle of incidence  increases;
\item and the sensitivity  to changes in refractive index of the infiltrating fluid decreases.
\end{itemize}

As well as directly shedding further light on SPP-based CTF sensors,
these results also highlight the  importance of using higher-order
homogenization methods when considering sculptured thin films
generally as platforms for  optical sensing. For example, in principle a
similar higher-order homogenization approach could be usefully employed for
chiral sculptured thin films, which are also promising candidates
for SPP-based optical sensing \c{ML_IEEE_SJ}.

\vspace{10mm}

\noindent {\bf Acknowledgments:} SSJ is fully sponsored by
Universiti Tun Hussein Onn Malaysia.

\newpage

\begin{figure}[!ht]
\centering
\subfigure[]{\includegraphics[width=3in]{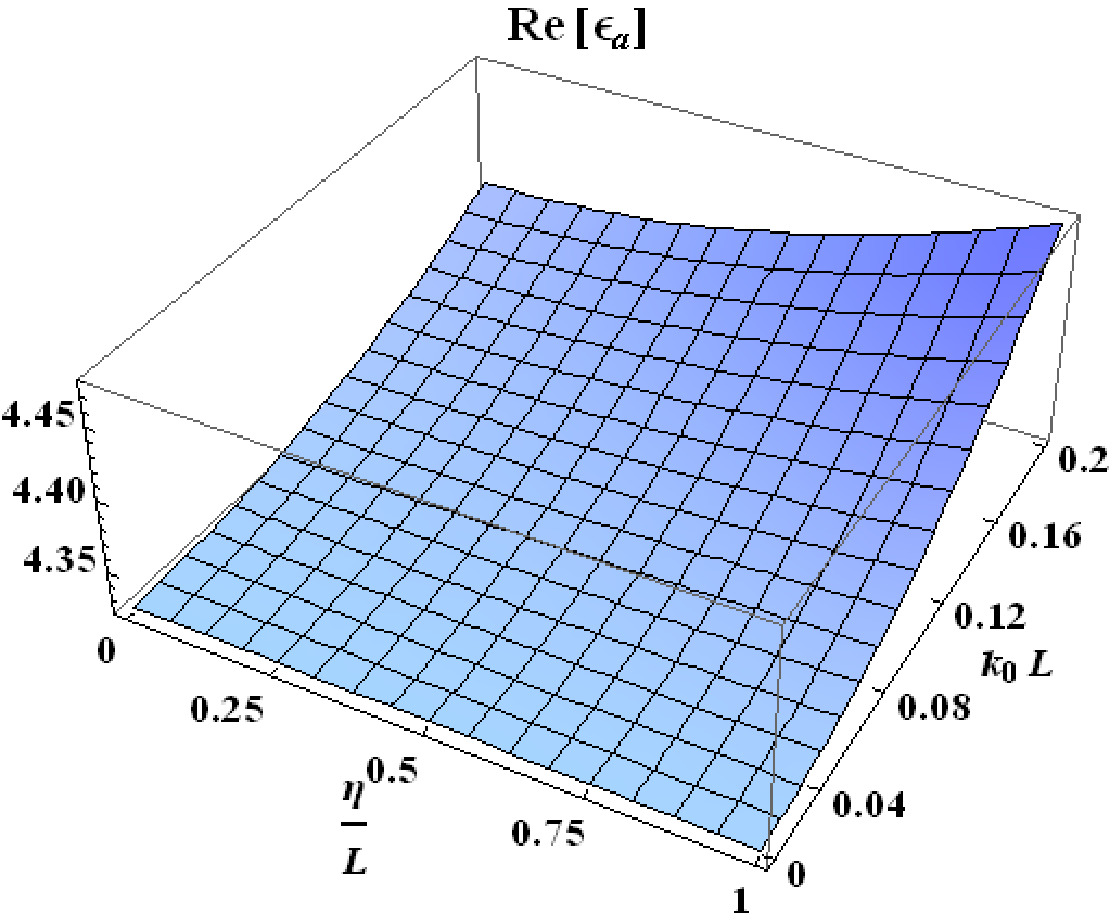}}
\subfigure[]{\includegraphics[width=3in]{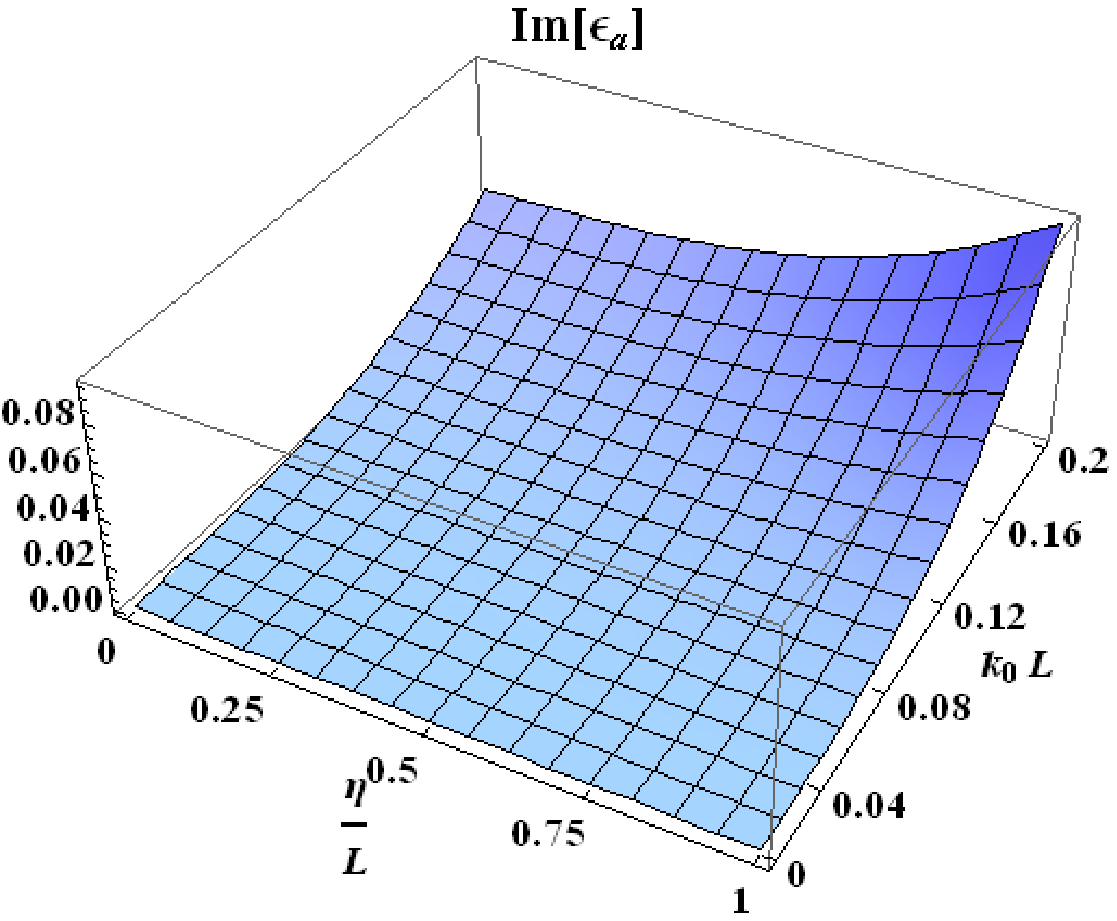}}
\subfigure[]{\includegraphics[width=3in]{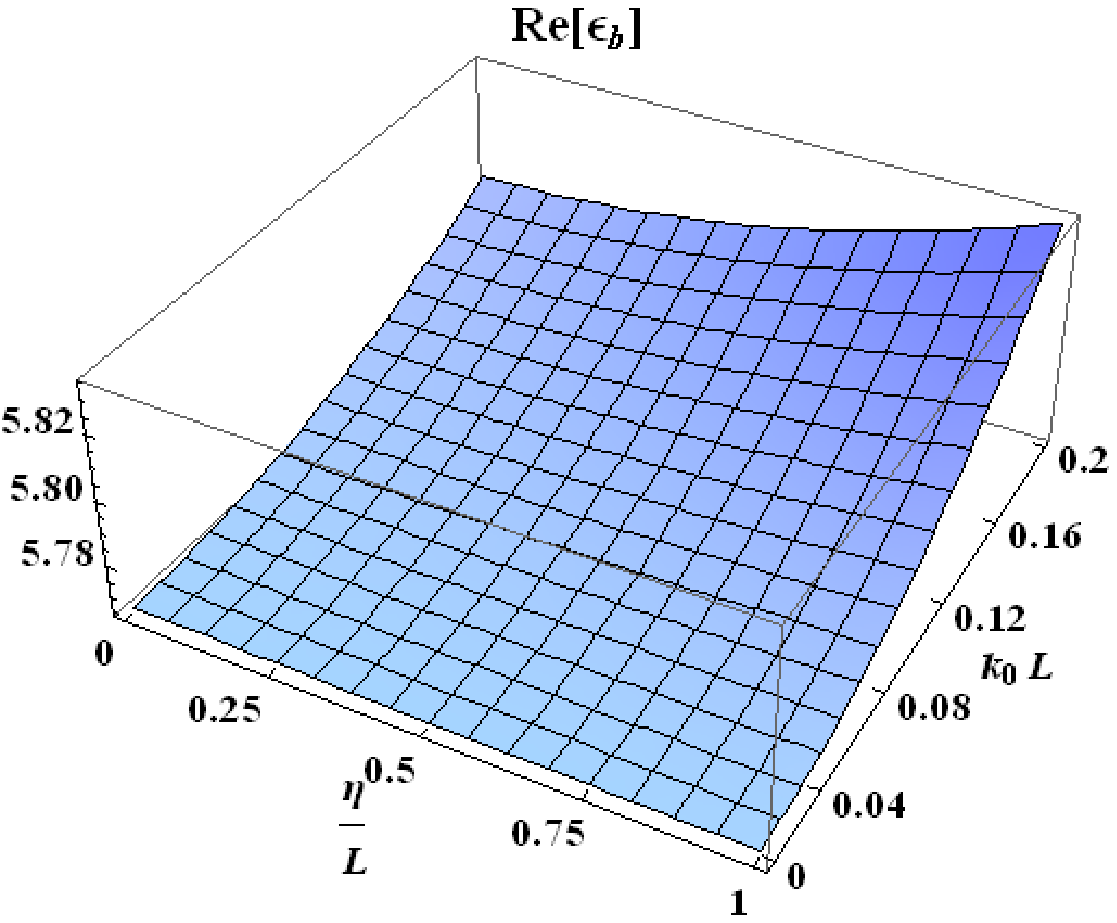}}
\subfigure[]{\includegraphics[width=3in]{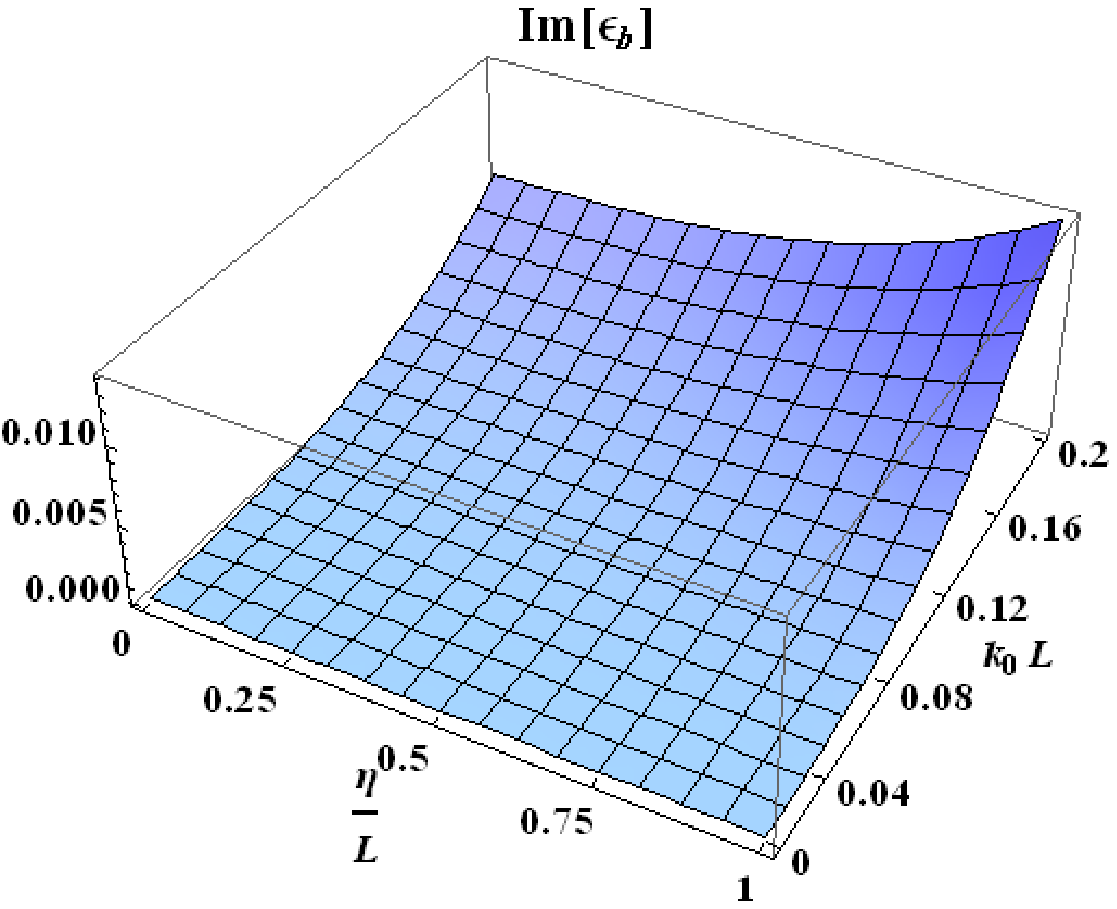}}
\subfigure[]{\includegraphics[width=3in]{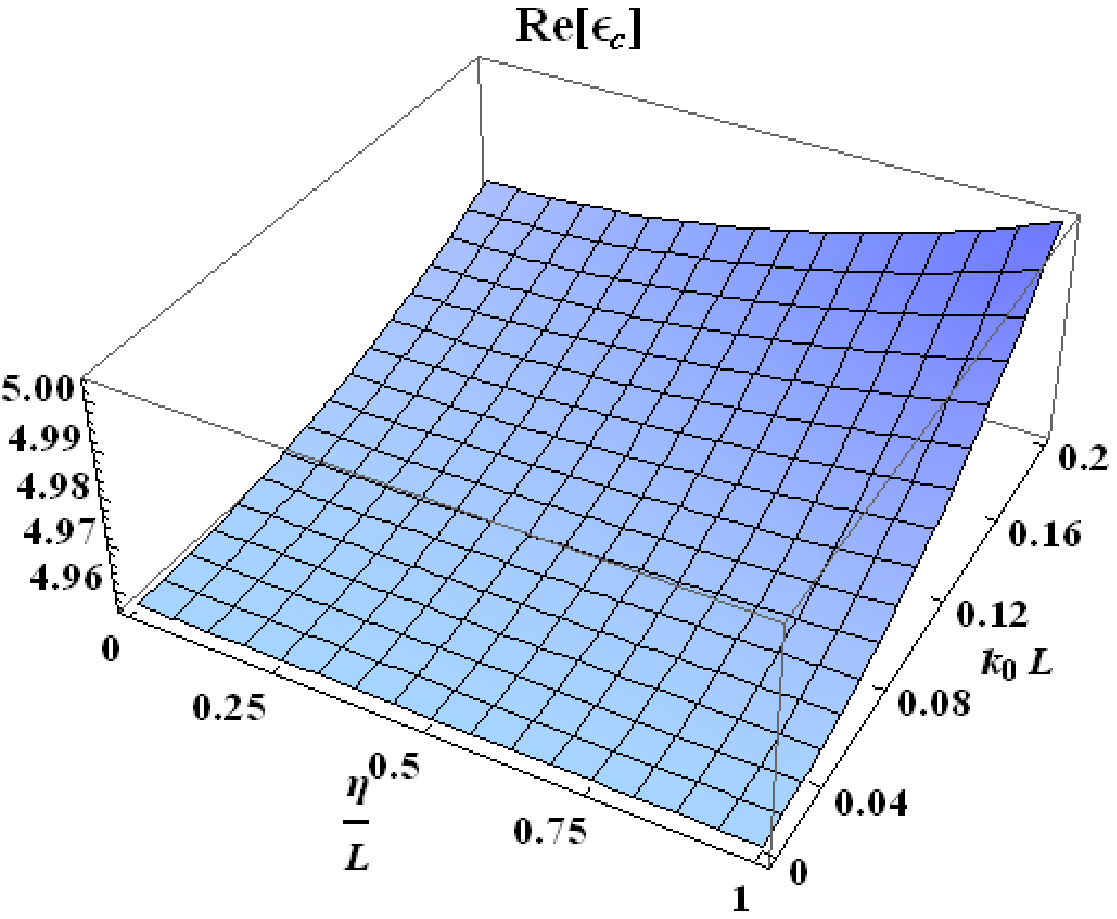}}
\subfigure[]{\includegraphics[width=3in]{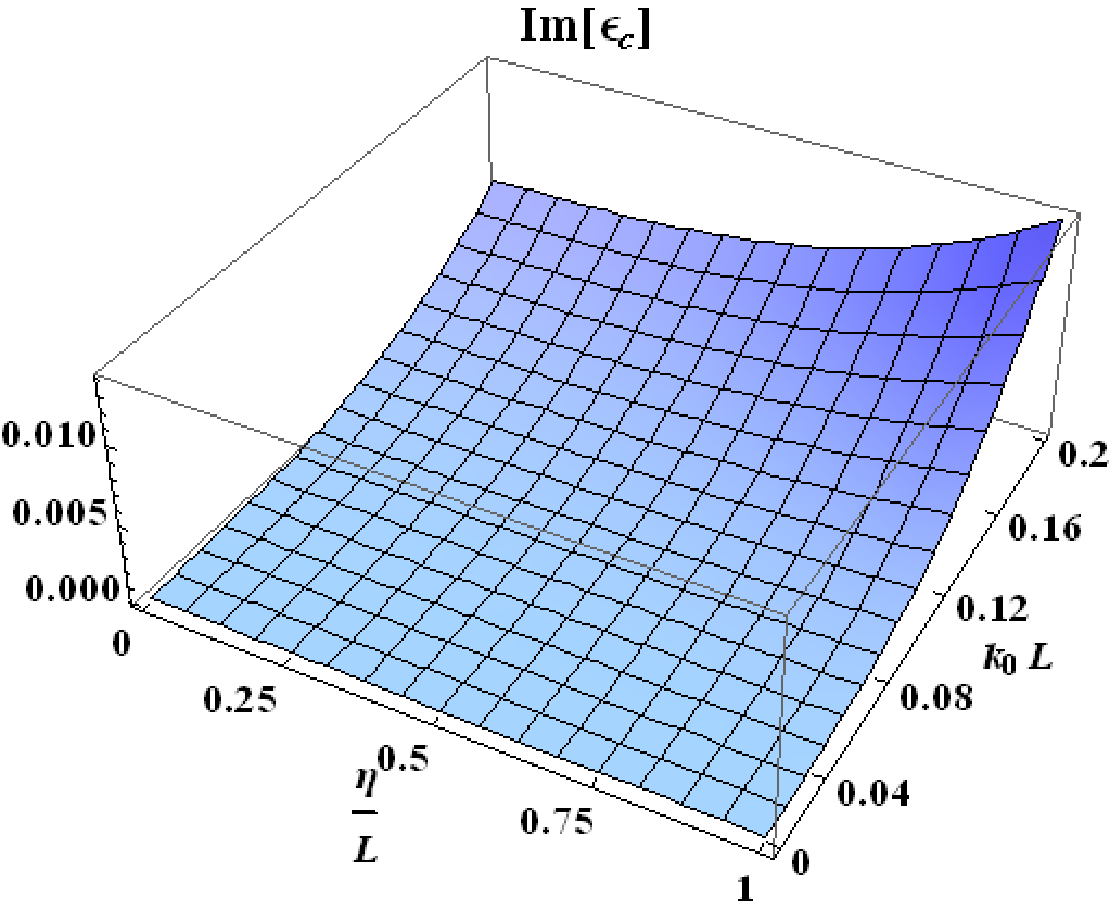}}
 \caption{ \l{Figure1}
 The relative permittivity parameters $\eps_{a2}$,  $\eps_{b2}$ and
 $\eps_{c2}$ of a fluid--infiltrated CTF, as computed using the extended second--order SPFT,
  plotted against $\eta/L \in (0,1)$ and $k_{0}L \in (0,0.2)$. The refractive index  of the fluid infiltrating
 the CTF is fixed at $n_\ell = 1.5$.}
\end{figure}

\newpage

\begin{figure}[!ht]
\centering \subfigure[]{\includegraphics[width=3in]{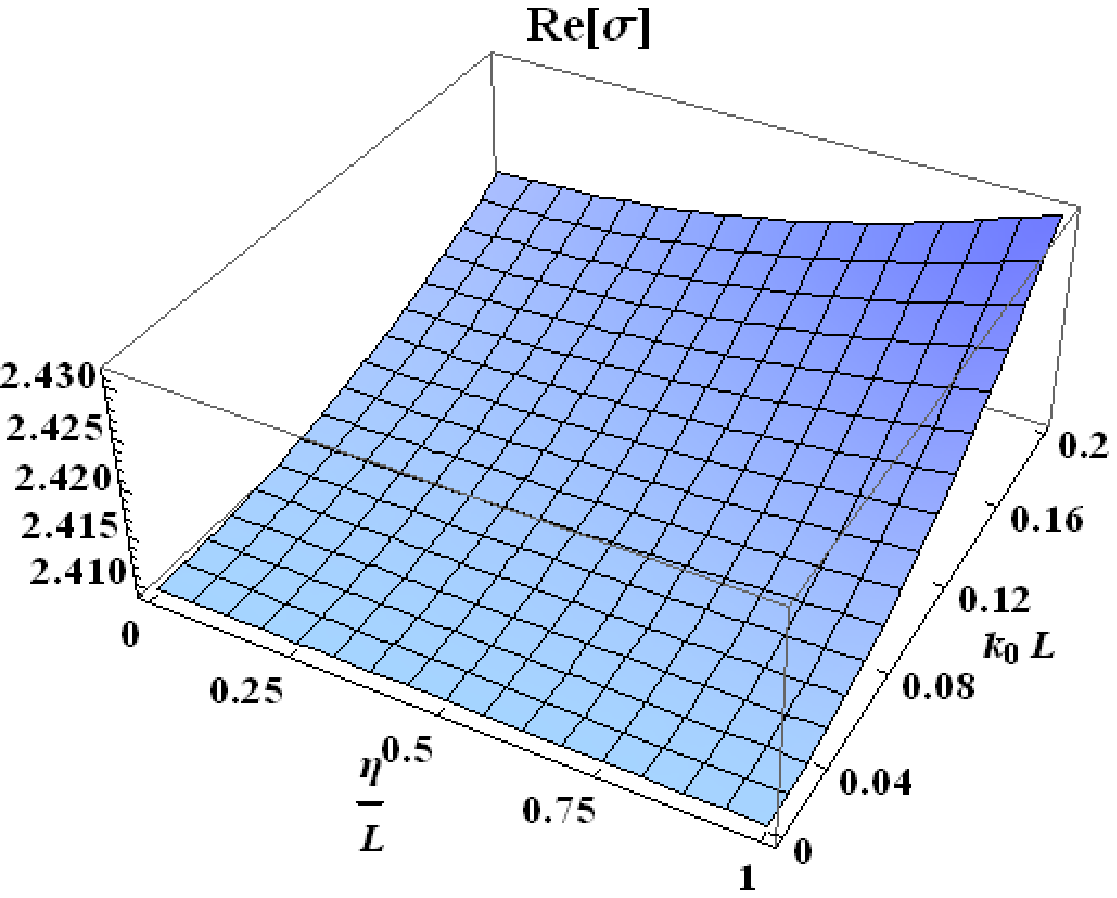}}
\subfigure[]{\includegraphics[width=3in]{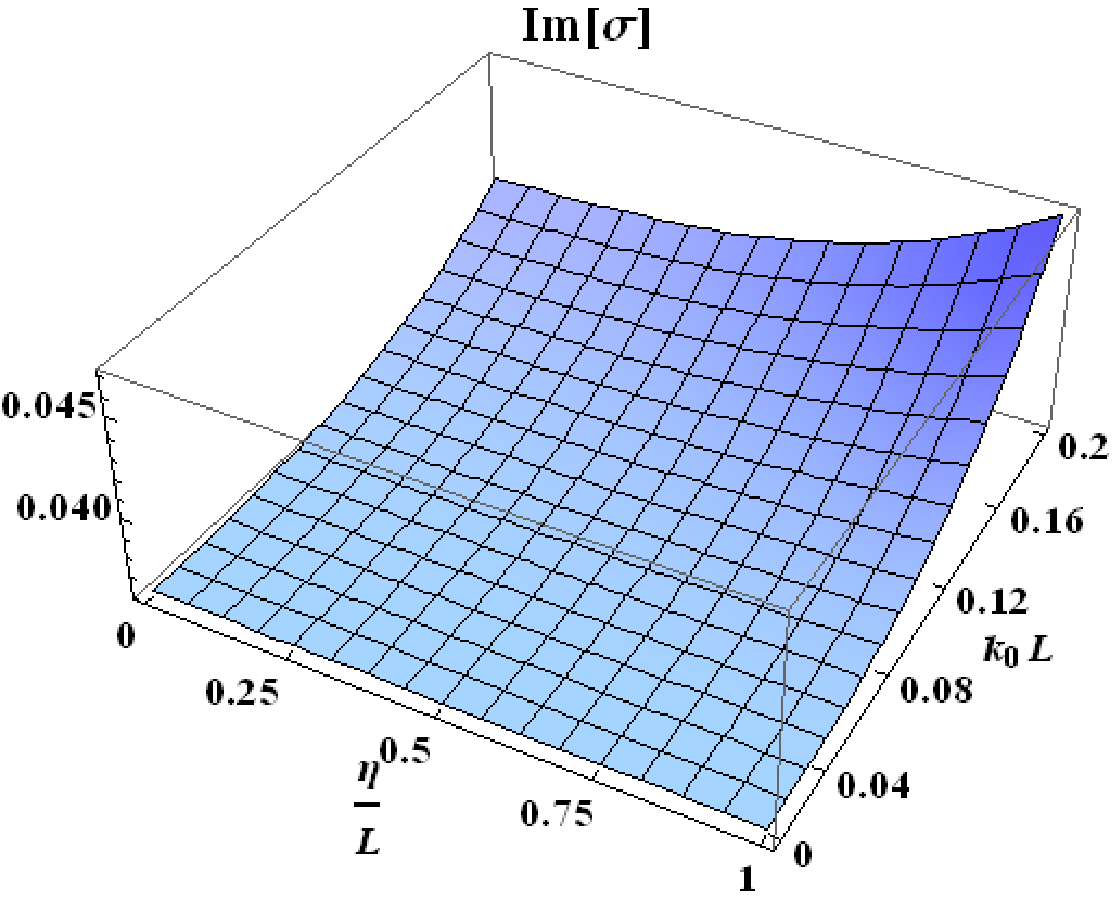}}
 \caption{ \l{Figure2}
 The real and imaginary parts of the relative wavenumber $\sigma$
 plotted against $\eta/L \in (0,1)$ and $k_{0}L \in (0,0.2)$.}
\end{figure}

\vspace{45mm}

\begin{figure}[!ht]
\centering
\includegraphics[width=3in]{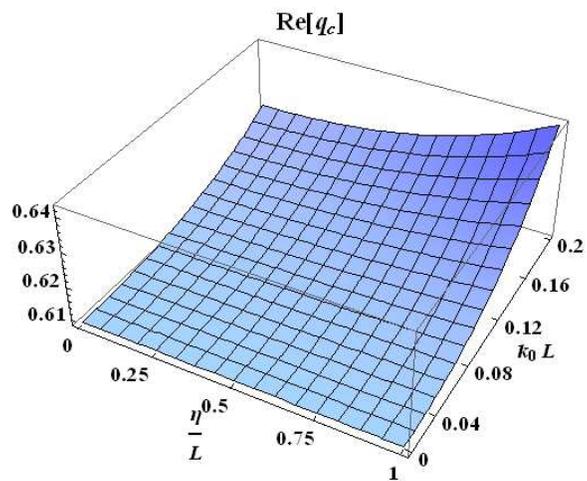}
 \caption{ \l{Figure3}
 The real part of $q_c$
 plotted against $\eta/L \in (0,1)$ and $k_{0}L \in (0,0.2)$.}
\end{figure}

\newpage

\begin{figure}[!ht]
\centering
\includegraphics[width=3in]{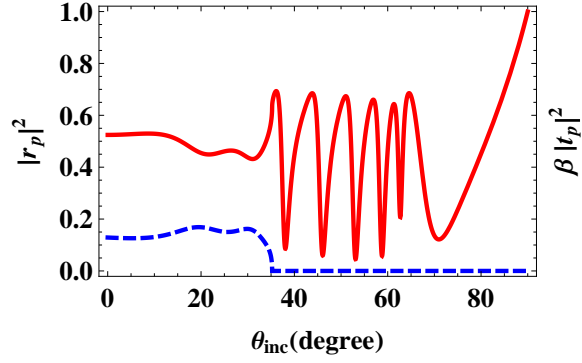}
 \caption{ \l{Figure4}
 The reflectance $\left| r_p \right|^2$ (red, solid curve) plotted against $\theta_{inc} $ (in degree), for the modified Kretschmann configuration with
   $\eps_{d} = 6.76$, $\eps_m =
 -56 + 21i$, $\eps_\ell = 2.25$,  $L_m = 10$ nm, $L_{\Sigma} = L_m + 1000$ nm,  $\eta/L = 1.0$ and $k_{0}L = 0.2$.
  Also plotted is the quantity $ \beta \left|
 t_p \right|^2$ (blue, dashed curve).}
\end{figure}

\vspace{45mm}

\begin{figure}[!ht]
\centering
\includegraphics[width=3in]{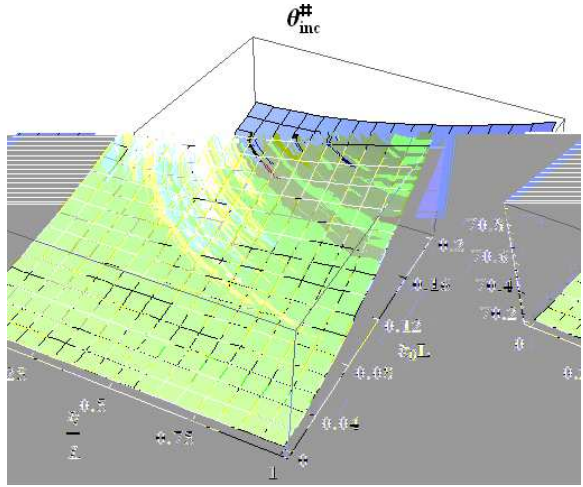}
 \caption{ \l{Figure6}
The angle  $\theta^{\sharp}_{inc}$ (in degree) for the scenario
represented in Fig.~\ref{Figure4},  plotted versus $\eta/L \in
(0,1)$ and $k_{0}L \in (0,0.2)$.}
\end{figure}

\newpage

\begin{figure}[!ht]
\centering
\includegraphics[width=3in]{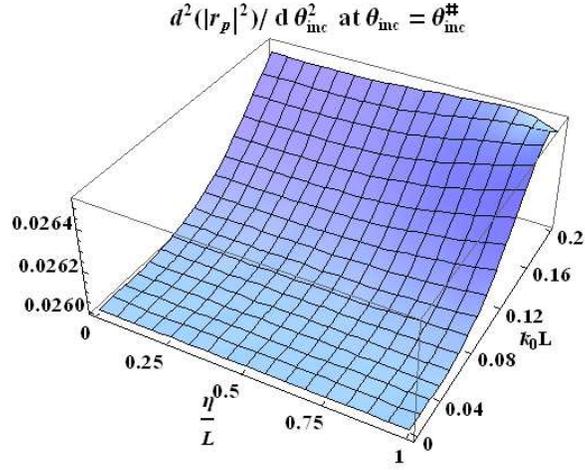}
 \caption{ \l{Figure7}
 As Fig.~\ref{Figure6} except that the quantity plotted is $d^2 \le \left| r_p
 \right|^2 \ri / d
 \theta^2_{inc}$   (in per degree per degree),  evaluated at $\theta_{inc} = \theta^{\sharp}_{inc}$.}
\end{figure}

\vspace{45mm}

\begin{figure}[!ht]
\centering
\includegraphics[width=3in]{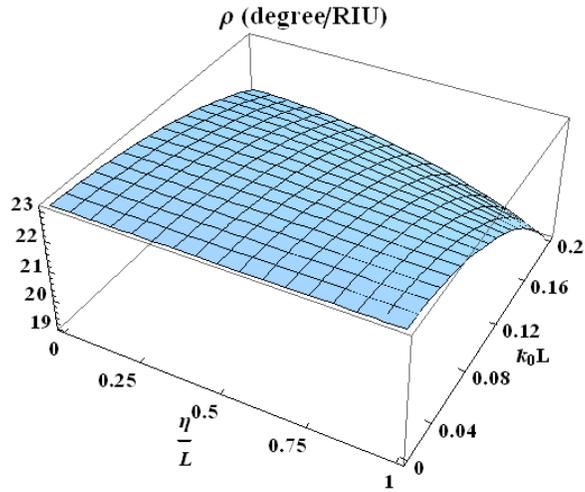}
 \caption{ \l{Figure8}
The figure of merit $\rho$ (in degree/RIU) plotted against $\eta/L
\in (0,1)$ and $k_{0}L \in (0,0.2)$ for $n_{\ell2} = 1.5$ and $n_{\ell1} = 1.0$.}
\end{figure}

\end{document}